\documentclass[12pt,aps,preprint,superscriptaddress,nofootinbib]{revtex4-1}

\usepackage{graphicx}
\usepackage{cancel}
\usepackage{amssymb}
\usepackage{textcomp}
\usepackage{amsmath}
\usepackage{bm}
\usepackage{times}
\usepackage{epsfig}
\usepackage{epstopdf}
\usepackage{color}
\usepackage{multirow}
\usepackage[colorlinks=true, pdfstartview=FitV, linkcolor=blue, citecolor=blue, urlcolor=blue]{hyperref}

\def\lsim{\mathrel{\raise.3ex\hbox{$<$\kern-.75em\lower1ex\hbox{$\sim$}}}}
\def\gsim{\mathrel{\raise.3ex\hbox{$>$\kern-.75em\lower1ex\hbox{$\sim$}}}}

\definecolor{orange}{rgb}{1,0.5,0}

\newcommand{\be}{\begin{equation}}
\newcommand{\ee}{\end{equation}}
\newcommand{\bea}{\begin{eqnarray}}
\newcommand{\eea}{\end{eqnarray}}
\def\tr{\mathrm{tr}}

\newcommand{\minigraph}[5][0.25in]{\begin{minipage}{#2}\begin{center}\includegraphics[width=#2]{#5}\\\vspace{#3}\hspace{#1}{\footnotesize #4}\end{center}\end{minipage}}

\begin{document}

\title{Type II Seesaw with scalar dark matter in light of AMS-02, DAMPE and Fermi-LAT}

\author{Tong Li}
\affiliation{
ARC Centre of Excellence for Particle Physics at the Terascale, School of Physics and Astronomy,
Monash University, Melbourne, Victoria 3800, Australia
}
\author{Nobuchika Okada}
\affiliation{
Department of Physics and Astronomy,
University of Alabama,
Tuscaloosa, AL 35487, USA
}
\author{Qaisar Shafi}
\affiliation{
Bartol Research Institute,
Department of Physics and Astronomy,
University of Delaware, Newark, DE 19716, USA
}

\begin{abstract}

The Standard Model (SM) supplemented by Type II Seesaw and a SM gauge-singlet scalar dark matter (DM)
  is a very simple framework to incorporate the observed neutrino oscillations and provide a plausible DM candidate.
In this framework, the scalar DM naturally has a leptophilic nature with a pair annihilating mainly into
  the SM SU(2)$_L$ triplet Higgs scalar of Type II Seesaw which, in turn, decay into leptons.
In this work, we consider indirect signatures of this leptophilic DM and examine
  the spectrum of the cosmic ray electron/positron flux from DM pair annihilations
  in the Galactic halo.
Given an astrophysical background spectrum of the cosmic ray electron/positron flux,
  we find that the contributions from DM annihilations can nicely fit the observed data
  from the AMS-02, DAMPE and Fermi-LAT collaborations,
  with a multi-TeV range of DM mass and a boost factor for the DM annihilation cross section
  of ${\cal O}(1000)$.
The boost factor has a slight tension with the Fermi-LAT data for gamma-ray
  from dwarf spheroidal galaxies, which can be ameliorated with an enhanced local DM density by a factor of about 2.

\end{abstract}


\maketitle

\section{Introduction}
\label{sec:Intro}

The existence of dark matter (DM) in the Universe has been established by a variety of
  cosmological and astronomical observations.
The nature of DM is still unknown and stands as one of the biggest mysteries in particle physics and cosmology.
The measurements of Galactic cosmic rays can provide indirect information of DM particles
  since their pair annihilations or late-time decays produce cosmic rays
  such as cosmic ray positrons and anti-protons.
As cosmic ray observations become more and more precise, a deviation (excess) of
  an observed cosmic ray flux from its astrophysical prediction can be used to learn about the nature of DM particles.
The latest and most precise measurements of the cosmic ray electron and positron (CRE) flux
  have been reported by the AMS-02~\cite{Aguilar:2014mma}, DAMPE~\cite{Ambrosi:2017wek}
  and Fermi-LAT~\cite{Abdollahi:2017nat} collaborations.
The increase of the positron spectrum above 100 GeV and the break of the CRE spectrum
  around 1 TeV are unexpected results from these measurements, possibly indicating an excess of the CREs originating from DM particles.

In this paper, we consider a very simple extension of the Standard Model (SM), which simultaneously resolves two major missing pieces in the SM, namely, a dark matter candidate and a neutrino mass matrix compatible with the observed neutrino oscillations.
This model was first proposed in Ref.~\cite{Gogoladze:2009gi} to account for an excess of cosmic ray positions
  reported by the PAMELA collaboration~\cite{Adriani:2008zr}.
After the first results of the AMS-02 \cite{Aguilar:2013qda}, detailed analysis for the cosmic ray position fluxes
  was performed in Ref.~\cite{Dev:2013hka}. This and similar models were also adopted to explain the line-like $e^++e^-$ excess around 1.4 TeV, recently reported by the DAMPE~\cite{Chen:2017tva,Li:2017tmd,Sui:2017qra}.
In the model, a SM gauge-singlet real scalar ($D$) is introduced along with an unbroken $Z_2$ symmetry.
The real scalar, which is a unique $Z_2$-odd field in the model, serves as the DM.
In addition to the scalar DM, a SM SU(2)$_L$ triplet scalar field ($\Delta$) is introduced to accommodate
   the observed neutrino oscillation phenomena \cite{Patrignani:2016xqp}
   via the Type II Seesaw mechanism \cite{Lazarides:1980nt, Mohapatra:1980yp, Magg:1980ut, Schechter:1980gr}.
Through a four-point interaction between the DM and triplet scalars,
   a pair of the DM particles mainly annihilates into a pair of the triplet scalars,
   whose subsequent decays generate leptonic final states as a new source of CREs in the Galactic halo.
The leptophilic nature of this DM particle is attributed to the Type II Seesaw mechanism.
The main purpose of this paper is to revisit the CRE spectrum for the Type II Seesaw model extended with scalar DM
   and identify a parameter region to fit the latest CRE spectrum measured by the AMS-02, DAMPE and Fermi-LAT collaborations.

This paper is organized as follows. In Sec.~\ref{sec:CR}, we describe the source and propagation of cosmic rays in the Galaxy including the DM contribution.
In Sec.~\ref{sec:Model} we discuss the Type II Seesaw model with scalar DM and its contribution to electron/positron cosmic rays. Our
numerical results of best-fit parameters in this model are given in Sec.~\ref{sec:Res}, and we summarize our conclusions in Sec.~\ref{sec:Con}.

\section{Electron/Positron Cosmic Rays in the Galaxy}
\label{sec:CR}

The key unknowns about cosmic rays in the Galaxy are their origin and propagation.
Primary cosmic rays originate from astrophysical processes, such as supernova explosions or pulsars. Their collisions
with intergalactic matter create secondary cosmic rays.
The propagation of cosmic rays can be described by the process of diffusion, in the form of the transport equation~\cite{Strong:2007nh}:
\begin{eqnarray}
{\partial \psi\over \partial t}&=&Q(\vec{r},p)+\vec{\nabla}\cdot \left(D_{xx}\vec{\nabla}\psi-\vec{V}\psi\right)+{\partial\over \partial p}p^2D_{pp}{\partial\over \partial p}{1\over p^2}\psi \nonumber\\
&&-{\partial\over \partial p}\left[\dot{p}\psi-{p\over 3}\left(\vec{\nabla}\cdot \vec{V}\right)\psi\right]-{\psi\over \tau_f}-{\psi\over \tau_r},
\label{propagation}
\end{eqnarray}
where $\psi(\vec{r},t,p)$ is the density of cosmic rays, $\vec{V}$ is the convection velocity,
  $\tau_f (\tau_r)$ is the time scale for fragmentation (radioactive decay), and $\dot{p}$ is the momentum loss rate.
In the above equation, the convection terms are governed by the Galactic wind and the diffusion in momentum space induces the re-acceleration process.
The spatial diffusion coefficient can then be written as
\begin{eqnarray}
D_{xx}=\beta D_0 (R/R_0)^\delta ,
\end{eqnarray}
with $R$ and $\beta$ being the rigidity and a particle velocity in unit of the speed of light, respectively.
This transport equation is numerically solved with given boundary conditions, that is,
  the cosmic ray density $\psi$ vanishes at the radius $R_h$ and the height $\pm L$ of
  the cylindrical diffusion zone in the Galactic halo.
The cosmic ray flux is then given by $\mathbf{v}\psi/(4\pi)$ with $\mathbf{v}$ being the cosmic ray velocity.

The source term in Eq.~(\ref{propagation}) for the primary cosmic rays can generally be given by the product of the spatial distribution
and the injection spectrum function:
\begin{eqnarray}
Q_i(\vec{r},p)=f(r,z) \, q(p) .
\end{eqnarray}
For the above spatial distribution, we use the following supernova remnants distribution,
\begin{eqnarray}
f(r,z)=f_0\left({r\over r_\odot}\right)^a{\rm exp}\left(-b \ {r-r_\odot\over r_\odot}\right){\rm exp}\left(-{|z|\over z_s}\right),
\label{snr}
\end{eqnarray}
where $r_\odot=8.5 \ {\rm kpc}$ is the distance between the Sun and the Galactic center,
and the height of the Galactic disk is taken to be $z_s=0.2 \ {\rm kpc}$.
The two parameters $a$ and $b$ are chosen to be 1.25 and 3.56, respectively~\cite{Trotta:2010mx}.
For the injection spectra of various nuclei and primary electrons, one can assume broken power law function with one or more breaks and indices.
These rigidity breaks and power law indices are the source injection parameters.

The secondary-to-primary ratio of cosmic ray nuclei (the Boron-to-Carbon ratio B/C) and the ratio of secondary isotopes
  (the Beryllium ratio $\rm ^{10}Be/^{9}Be$) are widely employed to constrain the cosmic ray propagation parameters~\cite{Lin:2014vja,Cuoco:2016eej,Cui:2016ppb,Feng:2016loc,Huang:2016tfo,Lin:2016ezz,Jin:2017iwg,Yuan:2017ozr,Niu:2017qfv},
   as they are respectively sensitive to the traveling path and the lifetime of cosmic rays in the Galaxy.
The source injection parameters of cosmic ray nuclei and electrons can be constrained by the measured proton flux data
   and the low energy regions of the CRE spectra, respectively.
Recently, the AMS-02 collaboration released abundant and precise data on the cosmic ray nuclei,
  e.g. proton~\cite{Aguilar:2015ooa}, B/C~\cite{Aguilar:2016vqr}, and Helium~\cite{Aguilar:2015ctt}.
Combining the latest CRE and nuclei data with old data sets such as from CREAM~\cite{Yoon:2011aa} and PAMELA~\cite{Adriani:2013as},
   one can constrain the propagation and source injection parameters in a statistical method.
Based on these well fitted parameters, an up-to-date primary electron and secondary electron/positron cosmic rays can be obtained with high precision.
Given this astrophysical background, at high energies, we are enabled to constrain extra compositions in cosmic rays
  such as annihilating dark matter which also produces electrons and positrons, in the light of the flux data newly reported
  by the AMS-02, DAMPE and Fermi-LAT collaborations.
In practice, we use the CRE background flux obtained in Ref.~\cite{Niu:2017hqe} (see also Ref.~\cite{Ge:2017tkd}) which takes into account the latest AMS-02 and DAMPE data based on the above global fitting procedure. We should keep in mind that both the explanation for the CRE excess measurements and the background modeling are the subject of debate. In addition to DM annihilation, there exist alternative astrophysical explanations for the difference between the data and prediction. The conclusion of the constraints on DM signal below is dependent on the background model adopted here.

The DM source term of the CREs for a self-conjugate DM particle can be described by the product of the spatial function and the spectrum distribution as follows
\begin{eqnarray}
Q(r,p)={1\over 2}{\rho^2(r)\over m_D^2}\langle \sigma v \rangle {dN\over dE}.
\label{DMQ}
\end{eqnarray}
Here, $\rho(r)$ is the DM spatial distribution, $\langle \sigma v \rangle$ is the total velocity-averaged DM annihilation cross section, and $dN/ dE$ is the energy spectrum of cosmic ray particles produced by the DM annihilation.
For the DM spatial distribution, we assume a generalized Navarro-Frenk-White (NFW) profile~\cite{Navarro:1995iw, Navarro:1996gj}
to describe DM halo within the Galaxy
\begin{eqnarray}
\rho(r)=\rho_s{(r/r_s)^{-\gamma}\over (1+r/r_s)^{3-\gamma}},
\end{eqnarray}
with $\gamma=1$ and $r_s=20$ kpc.
The coefficient $\rho_s$ is set to give the local DM density $\rho(r_\odot)=0.4 \ {\rm GeV/cm^3}$.
The spectrum $dN/dE$ depends on specific DM models.

\section{Type II Seesaw model with scalar DM}
\label{sec:Model}

\begin{table}[t]
\begin{center}
\begin{tabular}{|c|cc|c|}
\hline
           & SU(2)$_L$ & U(1)$_Y$ & $Z_2$  \\
\hline \hline
$ \ell_L^i   $ & {\bf 2}  & $-1/2$    & $+$  \\
$ H      $ & {\bf 2}  & $+1/2$    & $+$  \\
\hline
$ \Delta $ & {\bf 3}  & $+1$      & $+$  \\
\hline
$ D      $ & {\bf 1}  & $ 0  $    & $-$  \\
\hline
\end{tabular}
\end{center}
\caption{
Particle content of the Type II Seesaw model with scalar DM, relevant to our discussion in this paper.
In addition to the SM lepton doublets $\ell_L^i$ ($i=1,2,3$ being the generation index)
 and the Higgs doublet $H$,
 a complex SU(2)$_L$ triplet scalar $\Delta$ and a SM gauge-singlet real scalar $D$ are introduced,
 along with an unbroken $Z_2$ symmetry.
The triplet scaler $\Delta$ plays a key role in Type II Seesaw mechanism,
 while the scalar $D$ is the DM candidate.
}
\label{Tab:1}
\end{table}

In Table~\ref{Tab:1}, we list the particle content of the Type II Seesaw model with scalar DM, relevant for our discussion in this paper.
An odd parity under the unbroken $Z_2$ symmetry is assigned to the SM gauge-singlet scalar ($D$), which makes it stable and a suitable DM candidate.
The explicit form of the SU(2)$_L$ triplet scalar in terms of  three complex scalars
 (electric charge neutral ($\Delta^0$), singly charged ($\Delta^+$) and doubly charged ($\Delta^{++}$) scalars) is given by
\bea
 \Delta=\frac{\sigma^i}{\sqrt{2}}
   \Delta_i=\left(
 \begin{array}{cc}
    \Delta^+/\sqrt{2} & \Delta^{++}\\
    \Delta^0 & -\Delta^+/\sqrt{2}\\
 \end{array}\right) ,
\eea
where $\sigma^i$'s are Pauli matrices.

Following the notations of Ref.~\cite{Schmidt:2007nq}, the scalar potential relevant for Type II Seesaw is given by
\bea
 V(H, \Delta) &=&
 -m_H^2 (H^\dagger H)
  + \frac{\lambda}{2} (H^\dagger H)^2+M_\Delta^2 \, \tr \left[ \Delta^\dagger \Delta \right]
 + \frac{\lambda_1}{2} \left( \tr [\Delta^\dagger \Delta] \right)^2
 \nonumber\\
&+& \frac{\lambda_2}{2}\left(
 \left( \tr [\Delta^\dagger \Delta] \right)^2
 - \tr \left[  \Delta^\dagger \Delta \Delta^\dagger \Delta \right]
  \right)+\lambda_4 H^\dagger H \; \tr \left(\Delta^\dagger\Delta\right)
 + \lambda_5 H^\dagger
 \left[\Delta^\dagger, \Delta\right] H
\nonumber \\
&+& \left[ 2 \lambda_6 M_\Delta
    H^T i\sigma_2 \Delta^\dagger H +{\rm H.c.} \right],
 \label{H-Delta-Potential}
\eea
where the coupling constants $\lambda_i$
 are taken to be real without loss of generality.
The scalar potential relevant for DM $D$ is
\begin{eqnarray}
V(D)&=&{1\over 2}m_0^2D^2+\lambda_DD^4+\lambda_H D^2 H^\dagger H + \lambda_\Delta D^2 \tr(\Delta^\dagger \Delta).
\label{D-Potential}
\end{eqnarray}
Through the couplings $\lambda_H$
and $\lambda_\Delta$ in this scalar potential, a pair of scalar DM particles annihilates
into pairs of the Higgs doublet and the triplet, i.e. $DD\to H^\dagger H, \Delta^\dagger \Delta$.
A non-zero vacuum expectation value (VEV) of the Higgs doublet ($v$)
  generates a tadpole term for $\Delta$ through the last term in Eq.~(\ref{H-Delta-Potential}).
Minimizing the scalar potential, we obtain a non-zero VEV of the triplet Higgs as
  $\langle \Delta^0 \rangle = v_\Delta/\sqrt{2} \simeq \lambda_6 v^2/M_\Delta$,
  by which the lepton number is spontaneously broken.
In order to achieve the right scale of the electroweak symmetry breaking,
  the constraint, $v^2+v_\Delta^2=(246 \ {\rm GeV})^2$, must be satisfied.
After the electroweak symmetry breaking, one has the following triplet masses in the limit of $v\gg v_\Delta$,
\begin{eqnarray}
M_{\Delta^{\pm\pm}}^2=M_\Delta^2+{1\over 2}(\lambda_4+\lambda_5)v^2, \ M_{\Delta^{\pm}}^2=M_\Delta^2+{1\over 2}\lambda_4 v^2, \ M_{\Delta^{0}}^2=M_\Delta^2+{1\over 2}(\lambda_4-\lambda_5)v^2.
\end{eqnarray}
Here, $\Delta^0$ represents both CP-even and CP-odd neutral triplet Higgs bosons.
Taking a negative value for $\lambda_5$, we can arrange a non-degenerate spectrum for the triplet Higgs bosons
  with only the doubly charged Higgs boson being lighter than the scalar DM particle,
  i.e. $M_{\Delta^{\pm\pm}}<m_D<M_{\Delta^{\pm}},M_{\Delta^{0}}$.
In this case, through Eq.~(\ref{D-Potential}), a pair of scalar DM particles annihilates only into a pair of doubly charged Higgs bosons with a cross section
\begin{eqnarray}
\langle \sigma v\rangle(DD\to \Delta^{++}\Delta^{--})={1\over 8\pi m_D^2}\lambda_\Delta^2\sqrt{1-{M_{\Delta^{\pm\pm}}^2\over m_D^2}}.
\end{eqnarray}
Here we assume $\lambda_H\ll \lambda_\Delta$ to accommodate the direct DM detection~\cite{Li:2017tmd} and thus forbid annihilation into the Higgs doublet.
Since the annihilation modes into neutral and singly charged Higgs bosons are forbidden,
  neutrinos are not created from the prompt decay of the triplet scalar, thus evading a possible constraint from the IceCube experiment~\cite{Aartsen:2017kru,Zhao:2017nrt}.

As in the canonical Type II Seesaw, the triplet scalar ($\Delta$) has a Yukawa coupling with the lepton doublets given by
\bea
{\cal L}_\Delta &=&
 -\frac{1}{\sqrt{2}}\left(Y_\Delta\right)_{ij}
  \left(\ell_L^{i}\right)^T\, \mathrm{C} \, i\, \sigma_2\,  \Delta \, \ell_L^j +{\rm H.c.}\nonumber \\
  &=&
 -\frac{1}{\sqrt{2}} \left(Y_\Delta\right)_{ij} \,
  \left(\nu^{i}_L\right)^T \, \mathrm{C} \, \Delta^0 \, \nu_L^j+\frac{1}{2} \, \left(Y_\Delta\right)_{ij} \,
  \left(\nu^{i}_L\right)^T \, \mathrm{C} \, \Delta^+ \ e_L^j+\frac{1}{\sqrt{2}} \left(Y_\Delta\right)_{ij} \, \left(e^{i}_L\right)^T \,
  \mathrm{C} \, \Delta^{++} \  e_L^j  +  \text{H.c.}  ,\nonumber \\
\label{Yukawa}
\eea
where $\mathrm{C}$ is the charge conjugate matrix,
 and $\left(Y_\Delta\right)_{ij}$ denotes the elements
 of the Yukawa matrix.
In Eq.~(\ref{Yukawa}), the neutrino mass matrix is given by
\bea
  m_\nu = v_\Delta \, \left(Y_\Delta\right)_{ij} .
\eea
The Yukawa interactions of the doubly charged Higgs boson are
\begin{eqnarray}
{\cal L}_\Delta \supset
   -{1\over \sqrt{2}} (e_L^i)^T \ \mathrm{C} \ (Y_\Delta)_{ij} \ \Delta^{++} \ e_L^j,
\label{Y++}
\end{eqnarray}
and the partial decay width of the doubly charged Higgs boson into a same-sign dilepton is thus given by
\begin{eqnarray}
\Gamma(\Delta^{++}\to i^+ j^+)={1\over 8\pi(1+\delta_{ij})}|(Y_{\Delta})_{ij}|^2 M_{\Delta^{++}}, \ \ \ i,j = e,\mu,\tau.
\label{width}
\end{eqnarray}
For $v_{\Delta}\lesssim 10^{-4}$ GeV,
the decays of doubly charged Higgs are dominated by the above leptonic channels~\cite{Perez:2008ha}.
With this assumption we focus only on the same-sign leptonic decay modes of $\Delta^{++}$ in the analysis below. Thus, the decay branching ratios are only governed by Eq.~(\ref{width}) and are independent of the triplet Higgs mass.
Note that the Yukawa coupling matrix $Y_{\Delta}$ has a direct relation with the neutrino oscillation data,
\begin{eqnarray}
  Y_\Delta = \frac{1}{v_\Delta} U_{PMNS}^*  \, D_\nu  \, U_{PMNS}^{\dagger},
\label{Y++_2}
\end{eqnarray}
with $U_{PMNS}$ being the Pontecorvo-Maki-Nakagawa-Sakata (PMNS) neutrino mixing matrix and
  $D_\nu$ is a diagonal mass eigenvalue matrix for the light neutrinos.
Using the neutrino oscillation data~\cite{Patrignani:2016xqp} and the above decay width formula,
  we can obtain the patterns of the charged lepton flavors produced by the doubly charged Higgs boson decay.
In the following numerical calculations, we consider two cases for the light neutrino mass spectrum,
  namely, the normal hierarchy (NH) and the inverted hierarchy (IH) to account for the undetermined neutrino mass ordering.
For the two benchmarks, the obtained branching ratios of the doubly charged Higgs boson decay into the leptonic final states
  are shown in Table~\ref{BR}~\cite{Li:2018jns}. Remarkably, the branching ratio of doubly charged Higgs decay into $e^\pm e^\pm$ in IH is 50 times greater than that in NH, while all remaining decay channels are comparable between these two mass patterns.

\begin{table}[tb]
\begin{center}
\begin{tabular}{|c|c|c|c|c|c|c|}
\hline
BR  & $ee$ & $e\mu$ & $e\tau$ & $\mu\mu$ & $\mu\tau$ & $\tau\tau$
\\ \hline
NH & 1\% & 2\% & 2\% & 30\%  & 35\% & 30\%
\\ \hline
IH & 50\% & 1\% & 1\% & 12\%  & 24\% & 12\%
\\ \hline
\end{tabular}
\end{center}
\caption{Benchmark decay branching ratios of doubly charged Higgs boson for NH and IH spectra.}
\label{BR}
\end{table}

For the 4-body CRE spectrum we consider, one has
\begin{eqnarray}
&&{dN\over dE} = \sum_{i,j}{\rm BR}(\Delta^{\pm\pm}\to i^\pm j^\pm)\left(\frac{d\bar{N}_i}{dE}+\frac{d\bar{N}_j}{dE}\right),
\label{4body}
\end{eqnarray}
where $i,j = e,\mu,\tau$.
The cosmic ray spectrum $d\bar{N}/dE$ in the lab frame is given by the spectrum from the triplet Higgs decay in its rest frame, denoted by $dN/dE_0$, after a Lorentz boost~\cite{Elor:2015tva, Elor:2015bho}. Namely,
\begin{eqnarray}
{d\bar{N}\over dE}&=&\int^{t_{1,\rm max}}_{t_{1,\rm min}}{dx_0\over x_0\sqrt{1-\epsilon^2}}{dN\over dE_0},
\end{eqnarray}
where
\begin{eqnarray}
t_{1,\rm max}&=&{\rm min}\left[1,{2x\over \epsilon^2}\left(1+\sqrt{1-\epsilon^2}\right)\right],\\
t_{1,\rm min}&=&{2x\over \epsilon^2}\left(1-\sqrt{1-\epsilon^2}\right),
\end{eqnarray}
with $\epsilon=M_{\Delta^{\pm\pm}}/m_D$ and $x=E/m_D\leq 0.5$. For the propagation parameters of dark matter, we assume the MED model~\cite{Delahaye:2007fr,Donato:2003xg} with $\delta=0.7$, $D_0=0.0112 \ {\rm kpc}^2/{\rm Myr}$, $L=4 \ {\rm kpc}$ and the convective velocity $V_C=12 \ {\rm km/s}$, and use MicrOMEGAs 5.0~\cite{MicrOMEGAs} to calculate the CRE spectrum from DM annihilations.

\section{Results}
\label{sec:Res}

The obtained cosmic ray fluxes, together with the experimental data points, are put into a composite likelihood function, defined as
\begin{eqnarray}
\chi^2= \sum_i {(f_i^{\rm th}-f_i^{\rm exp})^2\over \sigma_i^2},
\label{sum}
\end{eqnarray}
where $f_i^{\rm th}=f_i^{\rm DM}+f_i^{\rm bkg}$ are the theoretical predictions
  of the CRE flux from the DM annihilation ($f_i^{\rm DM}$) plus the astronomical background ($f_i^{\rm bkg}$),
  while $f_i^{\rm exp}$ are the corresponding central value of the experimental data.
In order to take into account, amongst others, the uncertainty related to the fixed propagation parameters,
   we stipulate a 10\% uncertainty of the theoretical predictions.
The theoretical and experimental uncertainties are then combined in quadrature to yield the $\sigma_i$.
The sum in Eq.~(\ref{sum}) runs over the AMS-02 $e^+$ flux data ($E>1$ GeV, 70 data points), the DAMPE $e^++e^-$ data
   (without the excess point at $E\simeq 1.4$ TeV, 37 data points)
   or the Fermi-LAT $e^++e^-$ data (High-Energy region with $E>42$ GeV, 27 data points). Note that, for the DAMPE data, we exclude the data point with energy of about 1.4 TeV, as the evidence for the knee-like feature of $e^++e^-$ is rather strong but the line-like signal has to be confirmed in the future~\cite{Fowlie:2017fya}.

\begin{figure}[th!]
\begin{center}
\minigraph{7.cm}{-0.05in}{(a)}{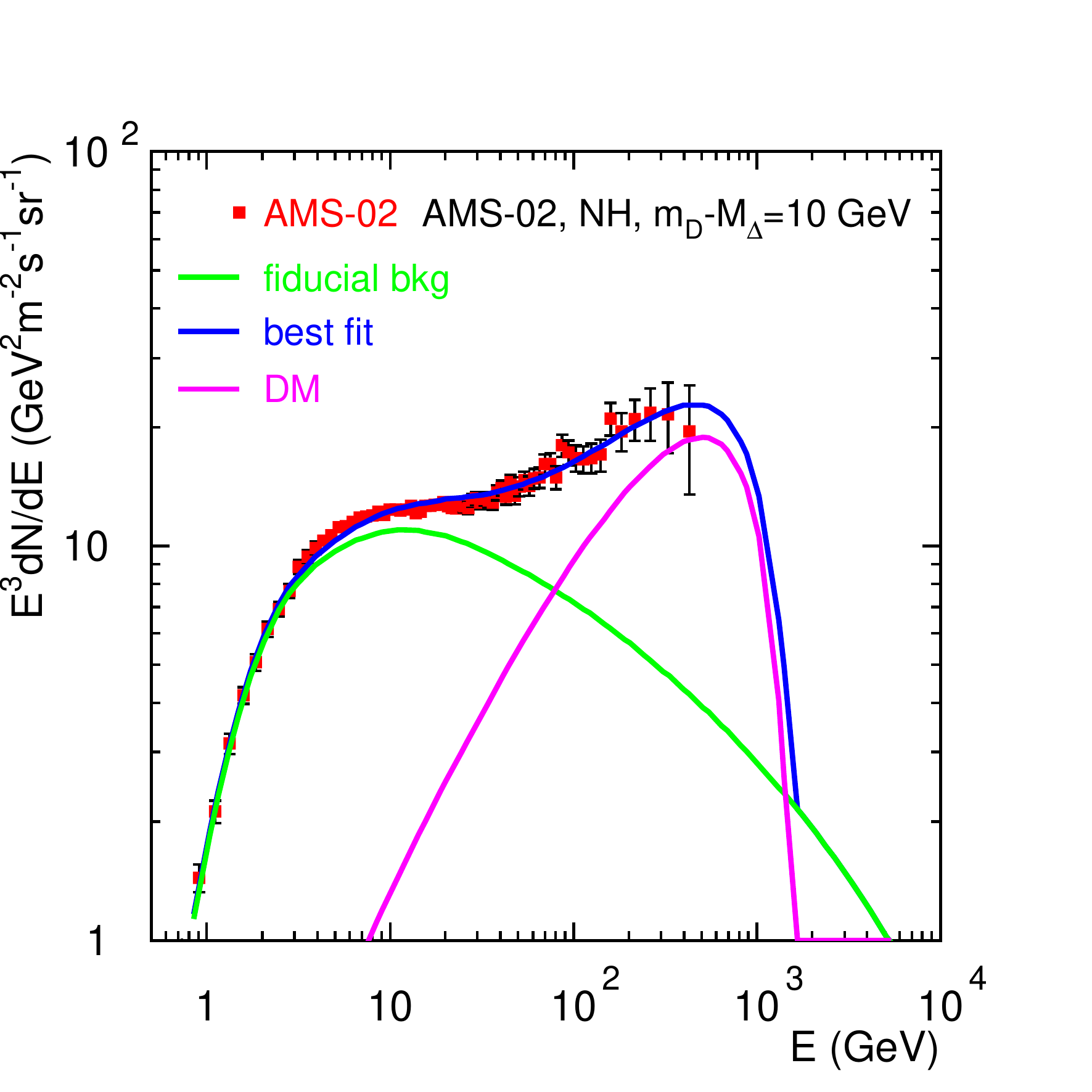}
\minigraph{7.cm}{-0.05in}{(b)}{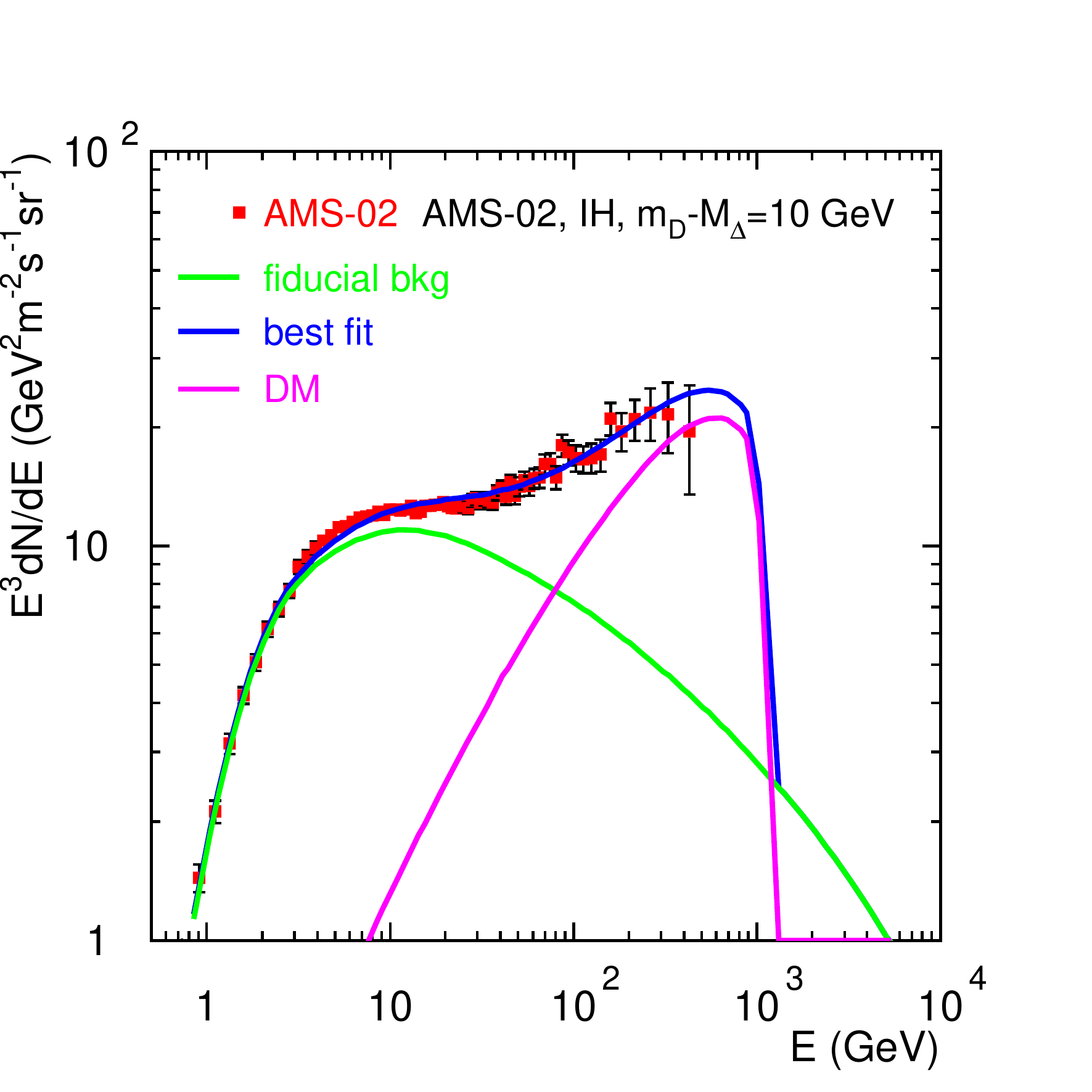}\\
\minigraph{6.5cm}{-0.05in}{(c)}{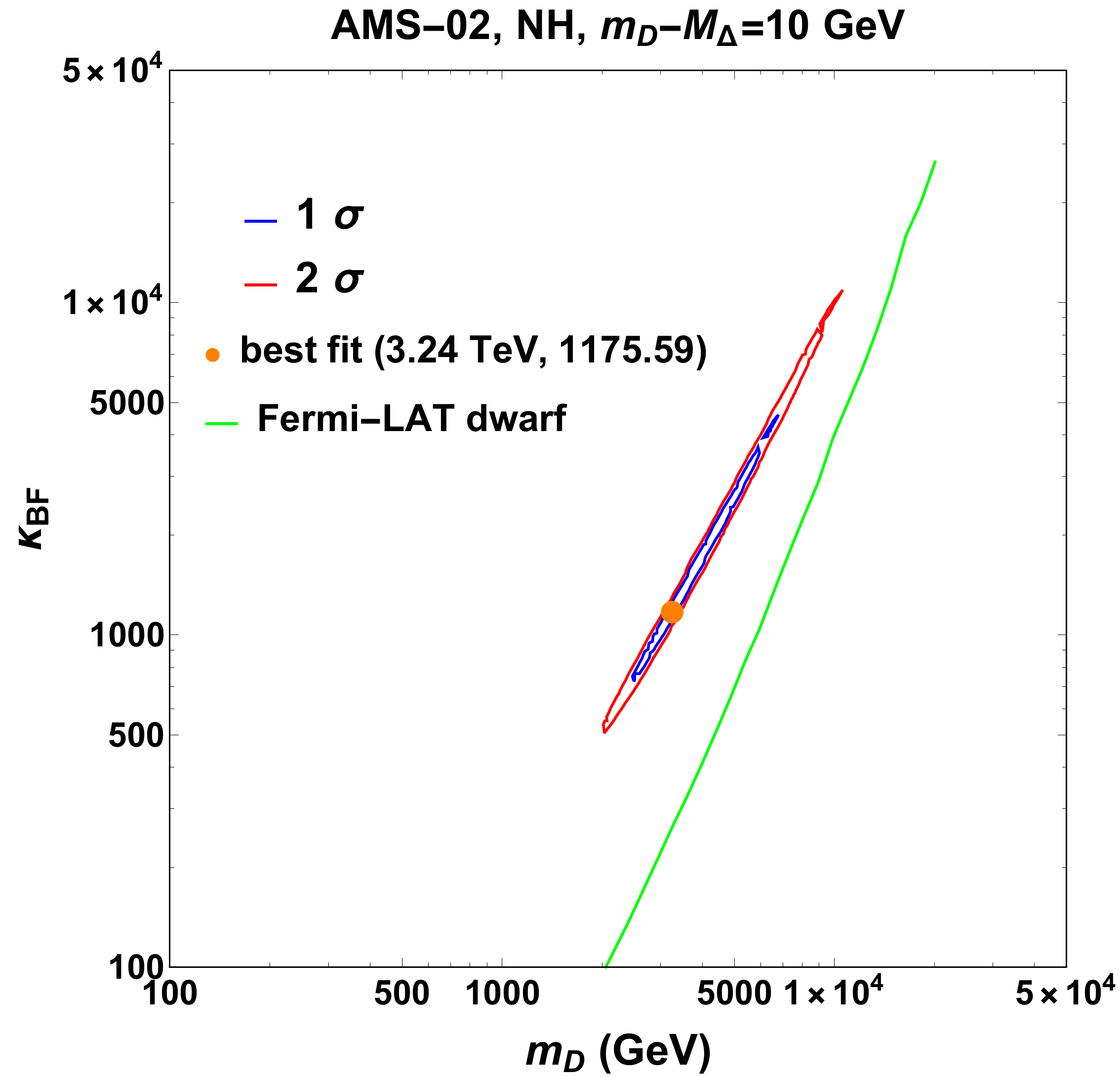}
\minigraph{6.5cm}{-0.05in}{(d)}{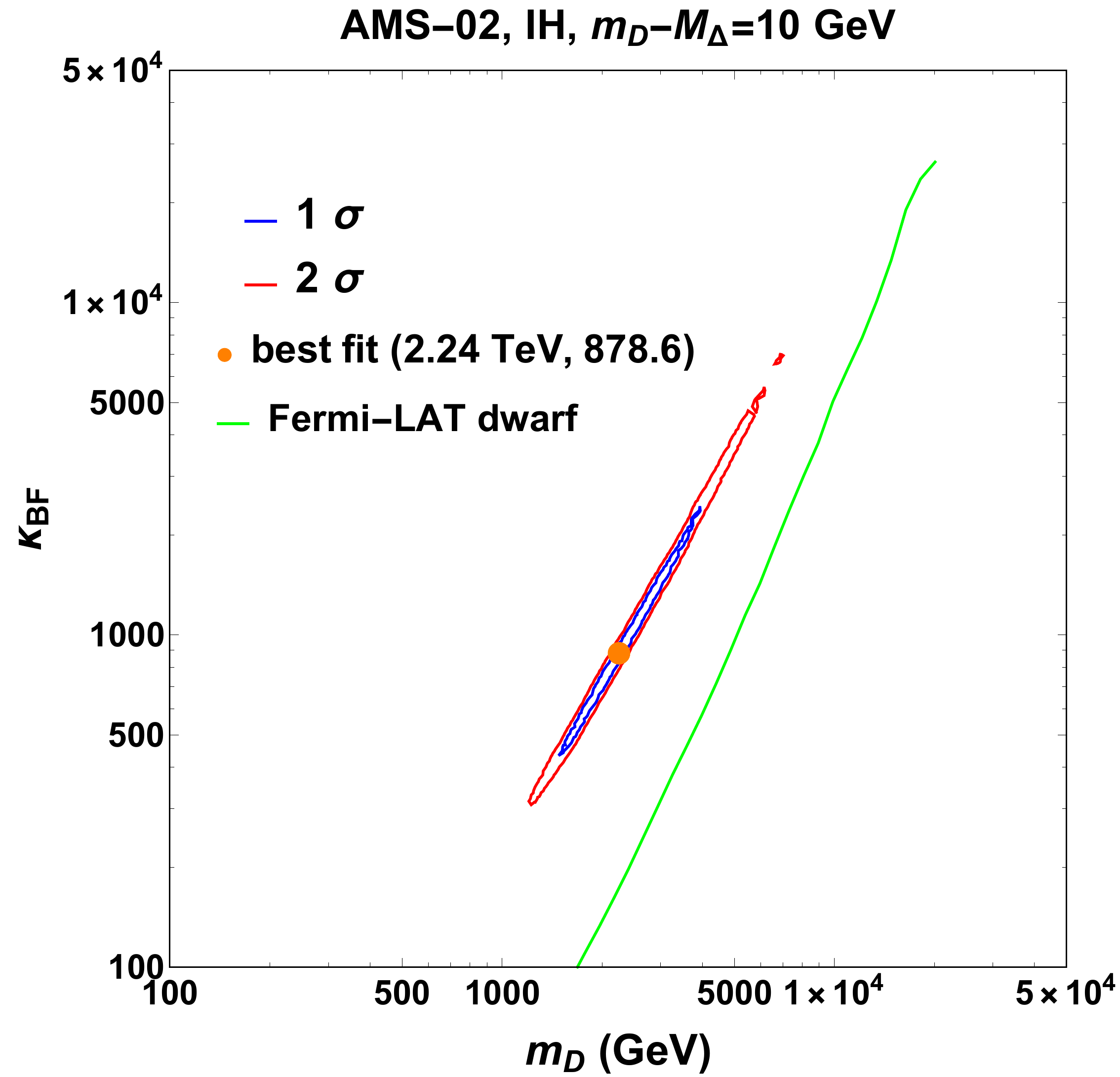}
\end{center}
\caption{(a) and (b): Comparison of the positron flux observed by the AMS-02 (red dots and dark error bars) with the Type II Seesaw model with scalar dark matter, for $m_D-M_{\Delta^{++}}=10$ GeV, and NH (a) and IH (b) spectra. The blue solid line shows the prediction of the total cosmic ray flux with dark matter parameter values that best fit the AMS-02
data. The total predicted flux is the sum of the background flux (green curve) and the dark matter
contribution (purple curve). (c) and (d): The AMS-02 favored region and best-fit point (orange point) of DM model parameters ($\kappa_{\rm BF}$ vs. $m_D$), for $m_D-M_{\Delta^{++}}=10$ GeV, and NH (c) and IH (d) spectra. The contours represent $1\sigma$ (blue) and $2\sigma$ (red) confidence regions. The upper limit on $\kappa_{\rm BF}$ by the Fermi-LAT is also shown in green curve.}
\label{AMS}
\end{figure}

\begin{figure}[th!]
\begin{center}
\minigraph{7.cm}{-0.05in}{(a)}{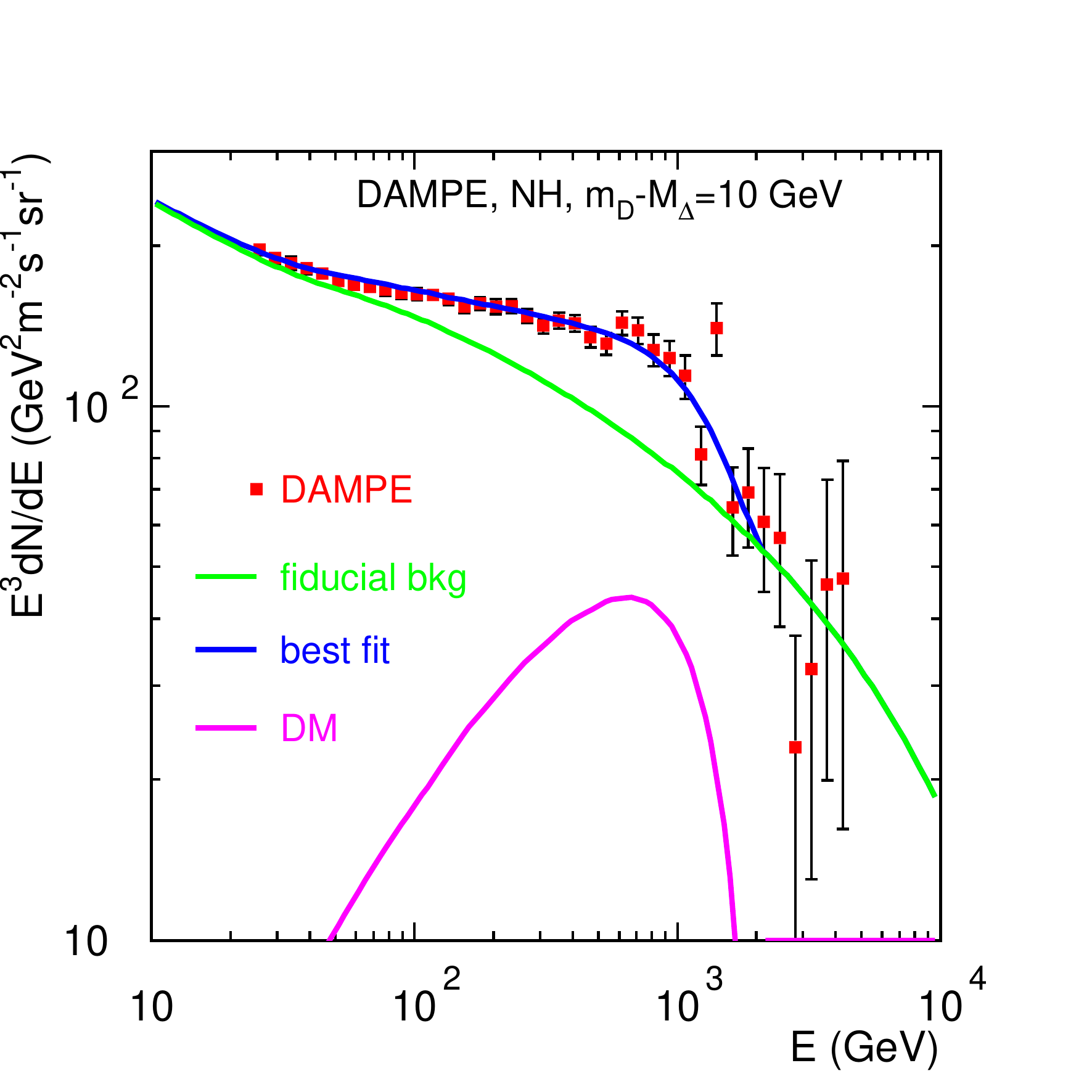}
\minigraph{7.cm}{-0.05in}{(b)}{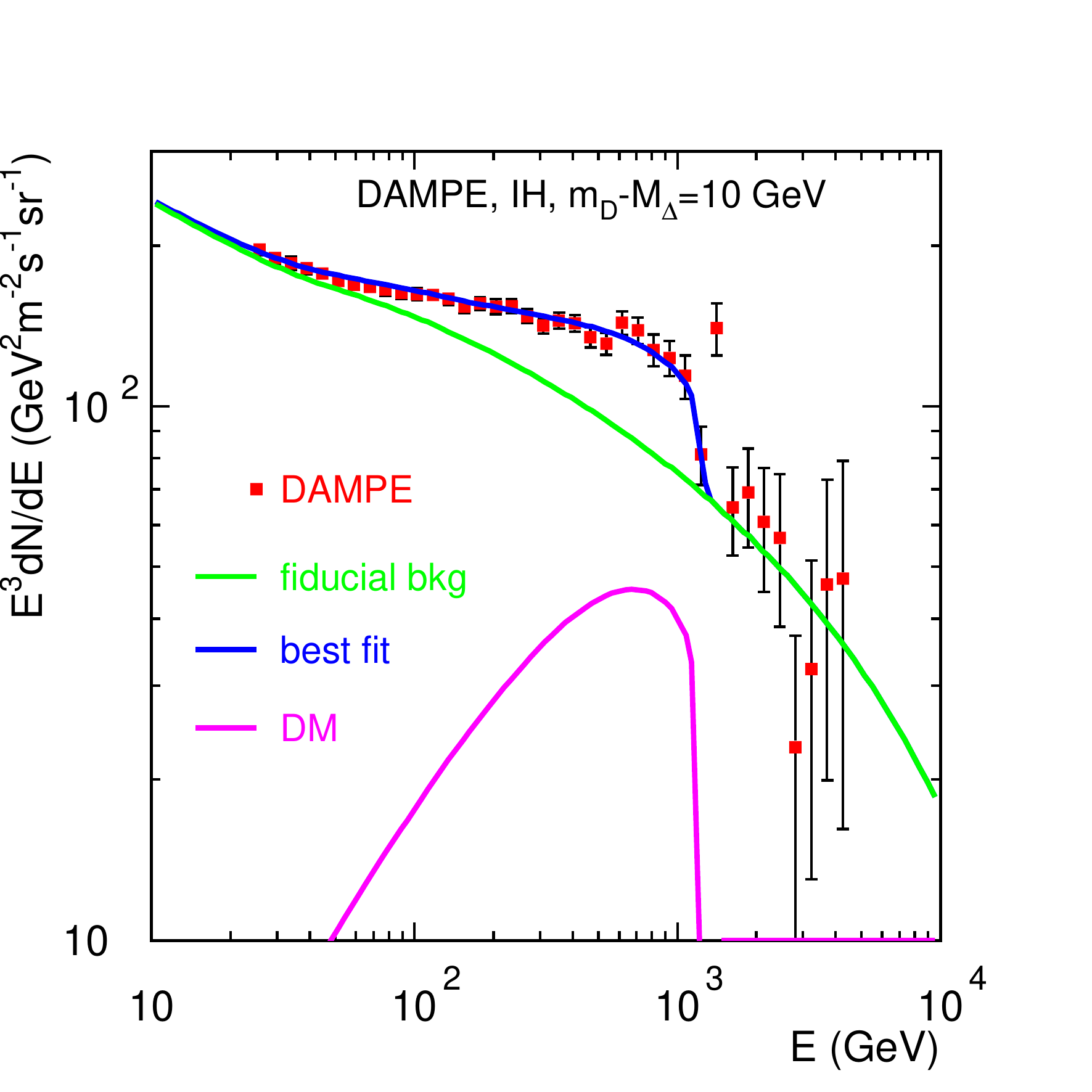}\\
\minigraph{6.5cm}{-0.05in}{(c)}{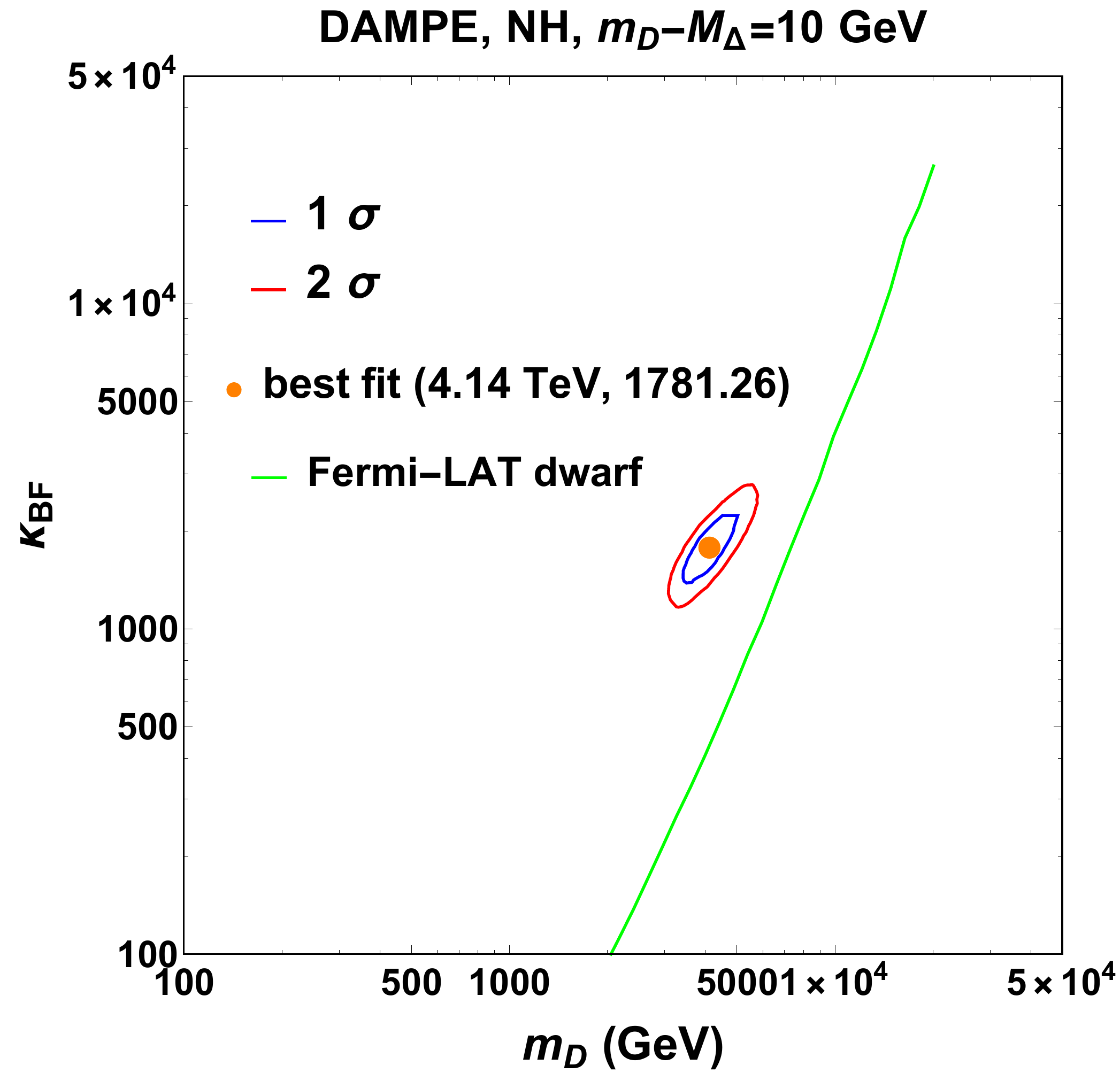}
\minigraph{6.5cm}{-0.05in}{(d)}{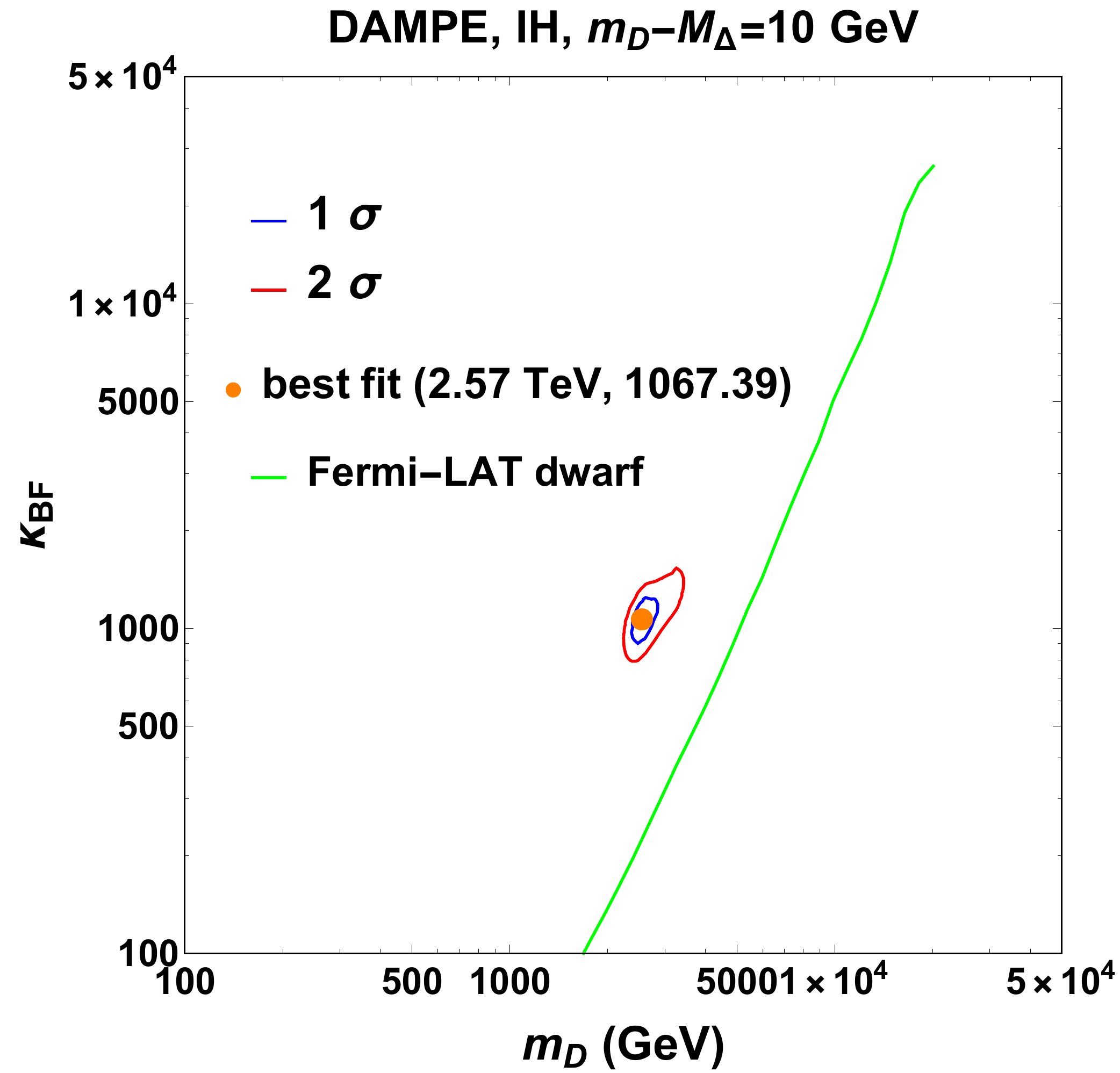}
\end{center}
\caption{
Electron plus positron flux that best fit the DAMPE data, as labeled in Fig.~\ref{AMS}.}
\label{DAMPE}
\end{figure}

\begin{figure}[th!]
\begin{center}
\minigraph{7.cm}{-0.05in}{(a)}{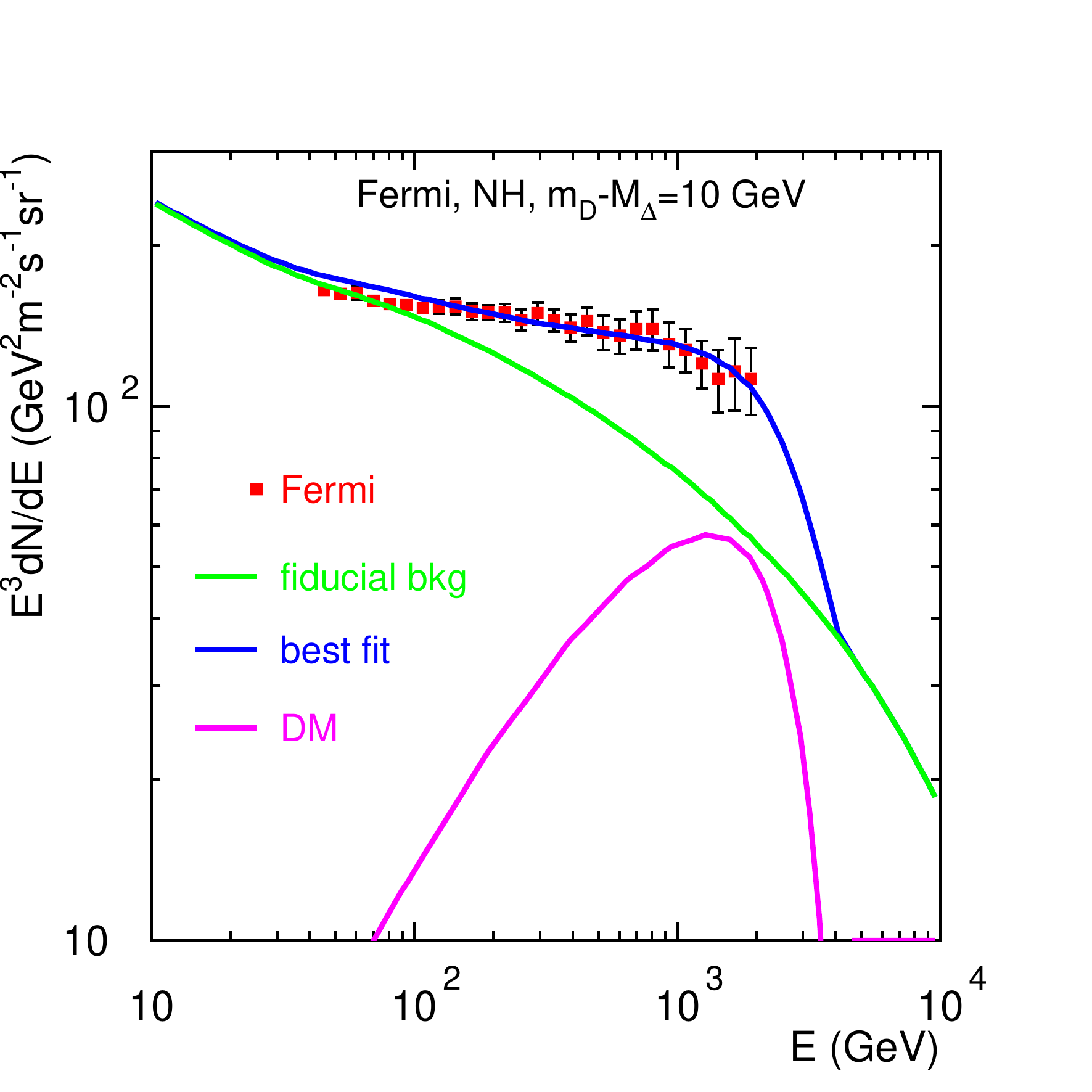}
\minigraph{7.cm}{-0.05in}{(b)}{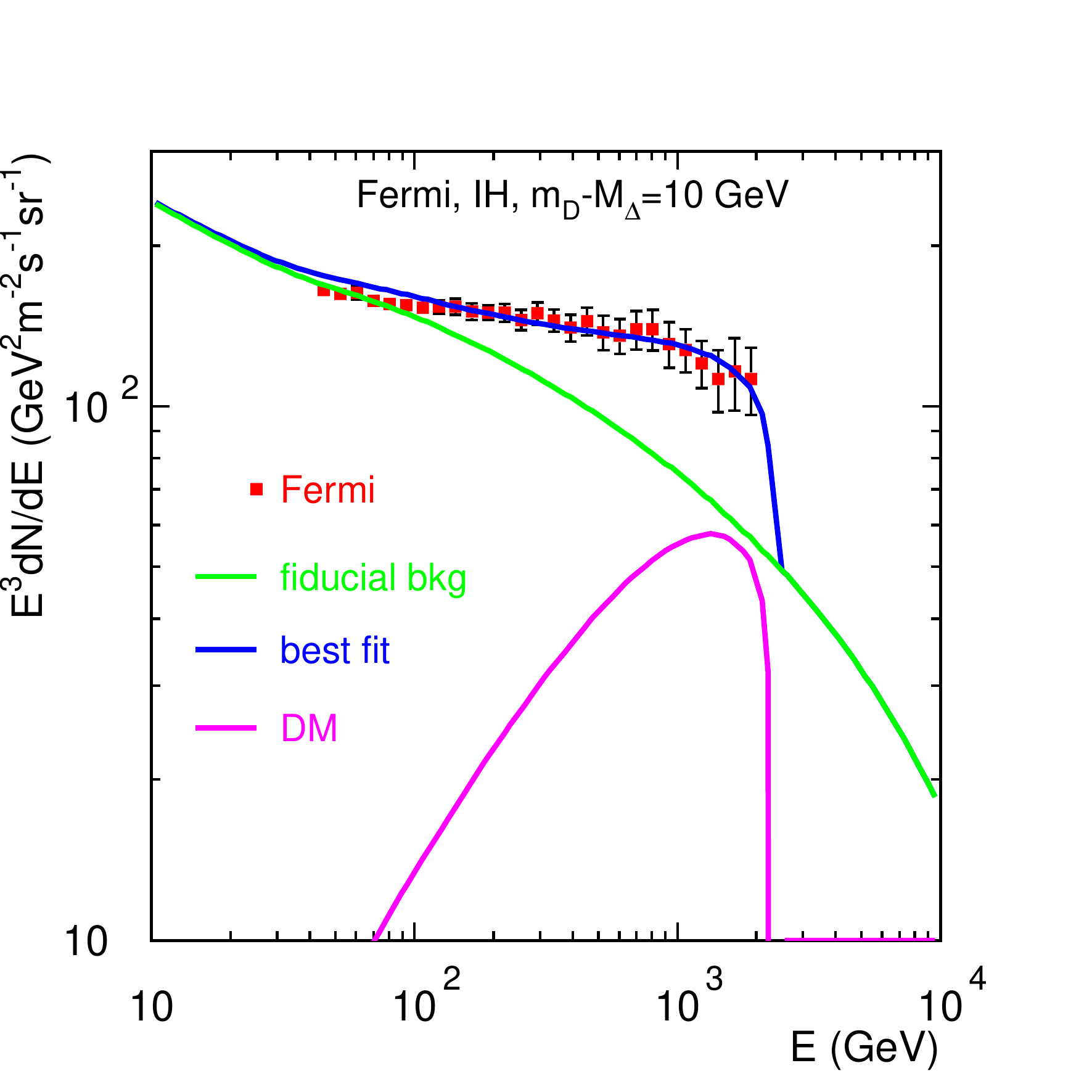}\\
\minigraph{6.5cm}{-0.05in}{(c)}{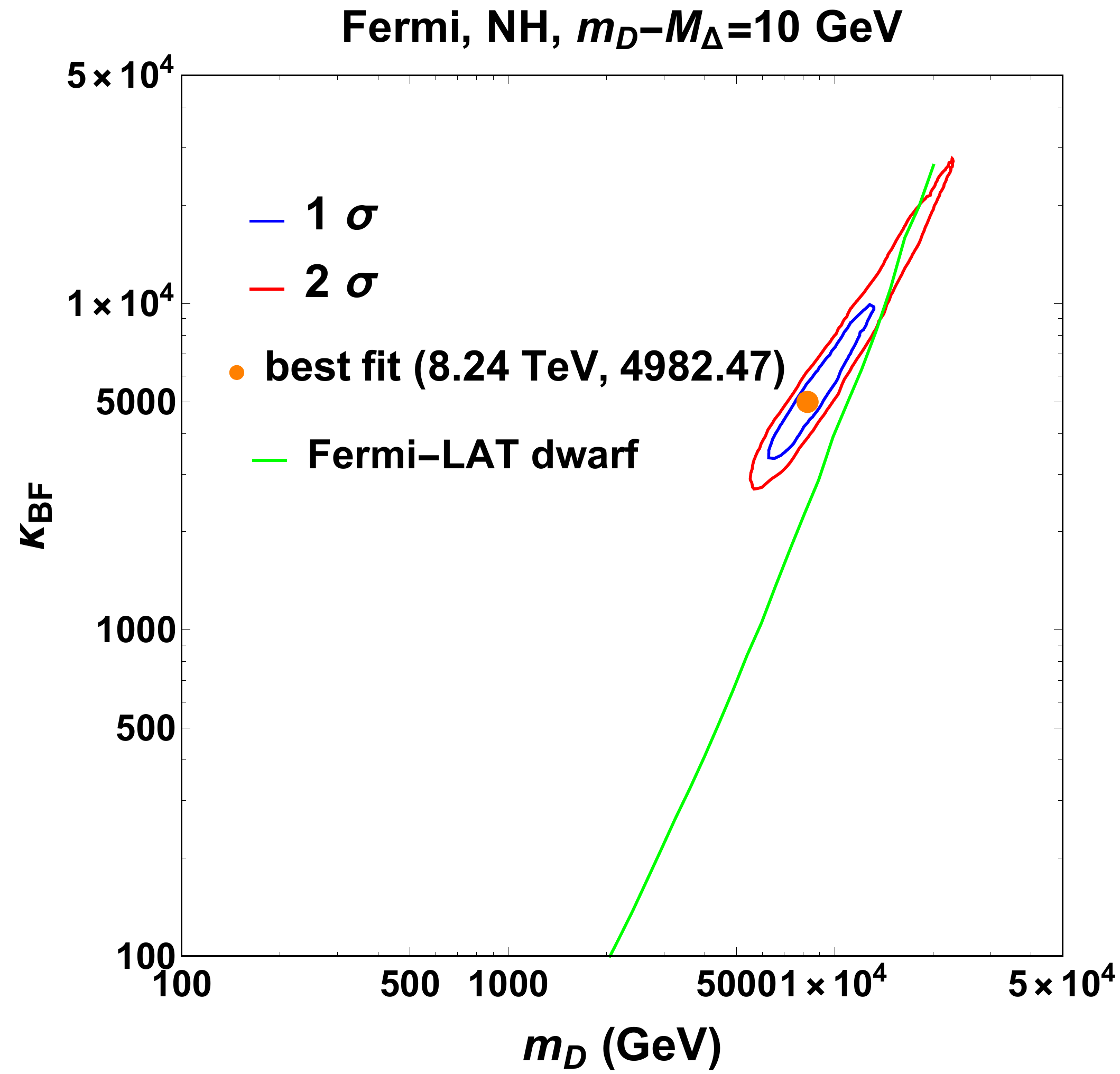}
\minigraph{6.5cm}{-0.05in}{(d)}{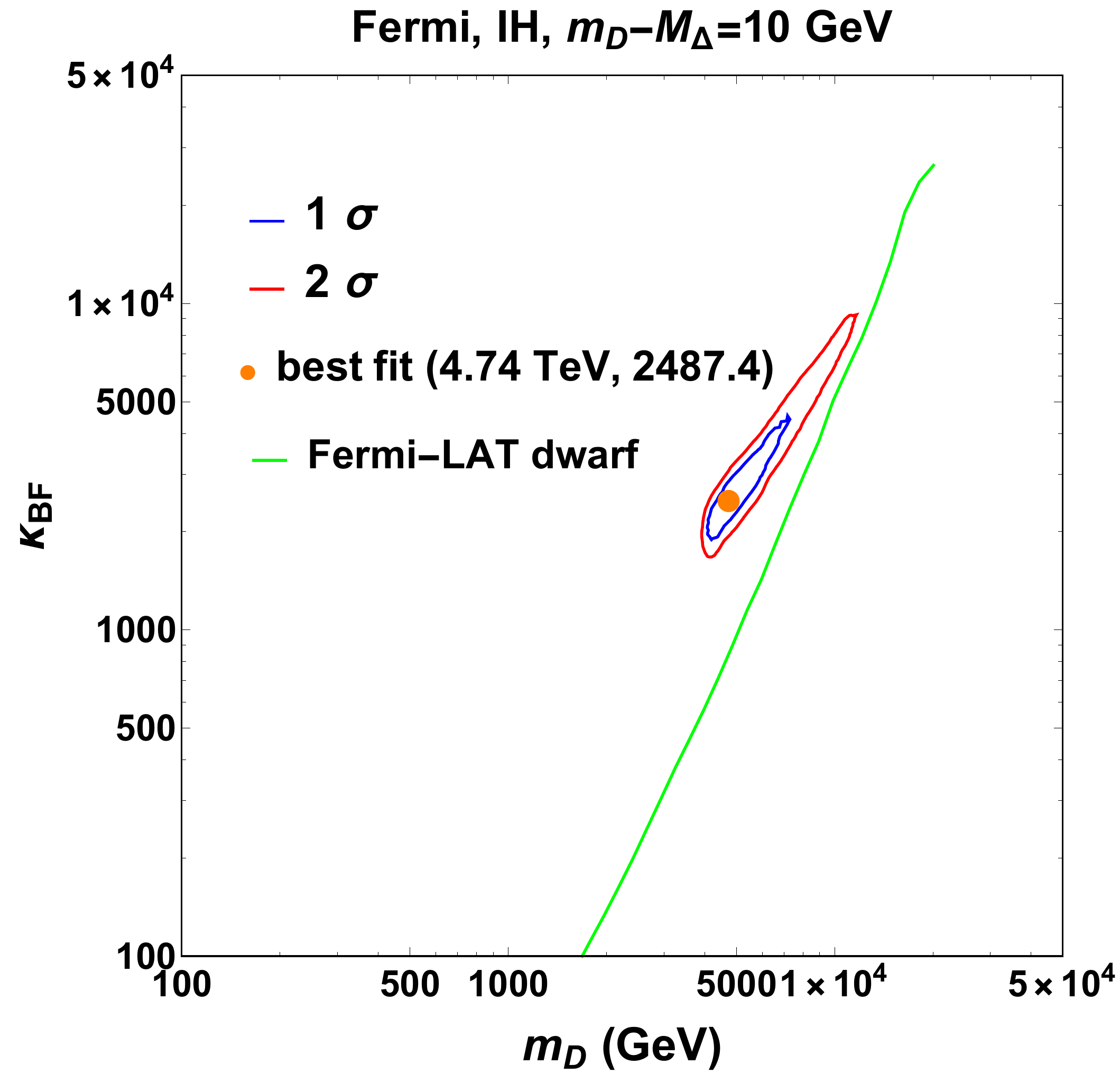}
\end{center}
\caption{Electron plus positron flux that best fit the Fermi-LAT data, as labeled in Fig.~\ref{AMS}.}
\label{Fermi}
\end{figure}

Due to the non-degenerate spectrum for triplet Higgs, we assume a mass benchmark with small difference $m_D-M_{\Delta^{++}}=10$ GeV,
  and $\langle \sigma v\rangle=\kappa_{\rm BF}\langle \sigma v\rangle_0$
  with $\langle \sigma v\rangle_0=3\times 10^{-26} \ {\rm cm}^3/{\rm s}$ (typical thermal DM annihilation cross section)
  and $\kappa_{\rm BF}$ being the boost factor.
Thus, we employ two free parameters, $m_D$ and $\kappa_{\rm BF}$,
  to fit the observed CRE spectrum.
In Figure~\ref{AMS} (a, b), we display our best-fit spectrum of cosmic ray positron
  in the Type II Seesaw with scalar DM, for $m_D-M_{\Delta^{++}}=10$ GeV, and NH and IH spectra,
  along with the AMS-02 data.
Figure~\ref{AMS} (c, d) also shows the favored region and the best-fit point of the DM model parameters
  in the plane of $\kappa_{\rm BF}$ vs. $m_D$ to fit the AMS-02 data.
In Figures~\ref{DAMPE} and \ref{Fermi}, we show our results corresponding to Figure~\ref{AMS}
  but for fitting the $e^++e^-$ flux measured by the DAMPE and the Fermi-LAT collaborations, respectively.
The corresponding favored regions and the best-fit points of the DM model parameters
  are also shown.
  Our results indicate that these CRE observations are consistent with
the dark matter hypothesis by improving the fit to the data.

Our results of best-fit parameters are summarized in Table~\ref{fitsum10}.
One can see that the increase of $e^+$ or $e^++e^-$ favors a multi-TeV DM mass and a boost factor $\kappa_{\rm BF} = \mathcal{O}(1000)$ for the annihilation cross section in this model. In particular, with 50 times greater branching ratio of doubly charged Higgs decay into $e^\pm e^\pm$, the favored DM mass in IH has a reduction of $30\%-40\%$ compared to that in NH. It is due to the sharper spectrum induced by the leading prompt $e^\pm e^\pm$ channel in IH. Moreover, as the first direct detection of $e^++e^-$ knee, DAMPE relatively prefers IH over NH, while both of NH and IH in this model indistinguishably fit AMS-02 and Fermi-LAT without a measured knee-like feature. More precise measurement of the flux for higher energy range in the future
should reveal a preference for NH or IH.

\begin{table}[tb]
\begin{center}
\begin{tabular}{|c|c|c|c|}
\hline
Best-fit $m_D, \kappa_{\rm BF}$  & AMS-02 & DAMPE & Fermi-LAT
\\ \hline
NH & 3.24 TeV, 1175.6 & 4.14 TeV, 1781.3 & 8.24 TeV, 4982.5 \\
   & $\chi^2_{\rm min}=9.9$ & $\chi^2_{\rm min}=7.2$ & $\chi^2_{\rm min}=2.73$
\\ \hline
IH & 2.24 TeV, 878.6 & 2.57 TeV, 1067.4 & 4.74 TeV, 2487.4 \\
   & $\chi^2_{\rm min}=10.0$ & $\chi^2_{\rm min}=5.53$ & $\chi^2_{\rm min}=2.83$
\\ \hline
\end{tabular}
\end{center}
\caption{Best-fit $m_D, \kappa_{\rm BF}$ and $\chi^2_{\rm min}$ for the NH and IH mass spectra, by AMS-02, DAMPE and Fermi-LAT, assuming $m_D-M_{\Delta^{++}}=10$ GeV.}
\label{fitsum10}
\end{table}

With a large amount of dark matter, dwarf galaxies serve as the bright targets for searching gamma rays from dark matter annihilation.
Since the Fermi-LAT experiments~\cite{Ackermann:2015zua, Drlica-Wagner:2015xua}
  have found no gamma ray excess from the dwarf spheroidal satellite galaxies (dSphs)
  of the Milky Way, following the Fermi's maximum likelihood analysis, one can place an upper limit on the DM annihilation cross section for a given $m_D$.
We perform the likelihood analysis and show the upper limit on $\kappa_{\rm BF}$
   by the Fermi-LAT dSphs in Figures~\ref{AMS}, \ref{DAMPE} and \ref{Fermi}. One can see a slight tension between the CRE favored region and the Fermi-LAT dSphs limit.
A slight enhancement by a factor of about 2 of the local DM density is needed to evade the Fermi-LAT dSphs constraint.

\section{Conclusion}
\label{sec:Con}
Type II Seesaw extension of the SM with a SM gauge-singlet scalar DM is a simple framework
  to supplement the SM with the desired neutrino mass matrix and a plausible DM candidate.
With a suitable choice of the model parameters, the scalar DM naturally becomes leptophilic;
  a pair of DM particles mainly annihilates into the doubly charged Higgs bosons which, in turn, decay into charged leptons.
We have calculated the spectrum of the cosmic ray electron/positron flux
  from DM pair annihilations in the Galactic halo.
Given an astrophysical background spectrum of the cosmic ray flux,
  we have found that the contributions from the DM annihilations can nicely fit
  the cosmic ray spectrum measured by the AMS-02, DAMPE and Fermi-LAT collaborations,
  with a multi-TeV range of DM mass and a boost factor for the DM annihilation cross section
  of ${\cal O}(1000)$.
Because of the Type II Seesaw structure for generating the neutrino mass matrix,
  the lepton flavor decomposition of the primary leptons from the doubly charged Higgs boson decay
  is determined by the pattern of the light neutrino mass spectrum and the neutrino oscillations data.
We have considered the NH and IH cases for the light neutrino mass spectrum.
As summarized in Table~\ref{fitsum10}, the IH case is preferred for fitting the DAMPE data, while
   both the NH and IH cases can equally fit the AMS-02 and the Fermi-LAT data.
We have also considered the Fermi-LAT constraint on the DM pair annihilation cross section
   and found a slight tension, which can be ameliorated with an enhanced local DM density
   by a factor of about 2.

\acknowledgments
This work is supported in part by the DOE Grant DE-SC0012447 (N.O.) and DE-SC0013880 (Q.S.).


\end{document}